\documentclass[aps,prb,twocolumn,floatfix,showpacs,superscriptaddress]{revtex4-1}

\usepackage{graphicx}
\usepackage{dcolumn}
\usepackage{amsmath}
\usepackage{bm}

\usepackage[breaklinks]{hyperref}
\usepackage[all]{hypcap}

\hypersetup{
    plainpages=false,
    bookmarks=false,         
    unicode=false,          
    pdfborder={0 0 0}
    pdftoolbar=true,        
    pdfmenubar=true,        
    pdffitwindow=false,     
    pdfstartview={FitH},    
    pdftitle={PRL},    
    pdfauthor={Vache Folle},     
    pdfproducer={LNCMI-CNRS-UGA-UPS-INSA}, 
    pdfkeywords={High} {Magnetic} {Fields}, 
    pdfnewwindow=true,      
    linktoc=section,
    colorlinks=true,       
    linkcolor=blue,          
    citecolor=red,        
    filecolor=magenta,      
    urlcolor=blue           
}

\begin{document}

\title{Ultrahigh magnetic field spectroscopy reveals the band structure of the 3D topological insulator Bi$_2$Se$_3$}

\author{A. \surname{Miyata}}
 \affiliation{Laboratoire National des Champs Magnetiques Intenses, CNRS-UGA-UPS-INSA, 143 Avenue de Rangueil, 31400 Toulouse, France}

 \author{Z. \surname{Yang}}
 \affiliation{Laboratoire National des Champs Magnetiques Intenses, CNRS-UGA-UPS-INSA, 143 Avenue de Rangueil, 31400 Toulouse, France}

 \author{A. \surname{Surrente}}
 \affiliation{Laboratoire National des Champs Magnetiques Intenses, CNRS-UGA-UPS-INSA, 143 Avenue de Rangueil, 31400 Toulouse, France}

 \author{O. \surname{Drachenko}}
 \affiliation{Laboratoire National des Champs Magnetiques Intenses, CNRS-UGA-UPS-INSA, 143 Avenue de Rangueil, 31400 Toulouse, France}

 \author{D. K. \surname{Maude}}
 \affiliation{Laboratoire National des Champs Magnetiques Intenses, CNRS-UGA-UPS-INSA, 143 Avenue de Rangueil, 31400 Toulouse, France}

 \author{O. \surname{Portugall}}
 \affiliation{Laboratoire National des Champs Magnetiques Intenses, CNRS-UGA-UPS-INSA, 143 Avenue de Rangueil, 31400 Toulouse, France}

 \author{L. B. \surname{Duffy}}
\affiliation{Clarendon Laboratory, University of Oxford, Parks Road, Oxford, OX1 3PU, UK}

\author{T. \surname{Hesjedal}}
\affiliation{Clarendon Laboratory, University of Oxford, Parks Road, Oxford, OX1 3PU, UK}

 \author{P. \surname{Plochocka}}
 \affiliation{Laboratoire National des Champs Magnetiques Intenses, CNRS-UGA-UPS-INSA, 143 Avenue de Rangueil, 31400 Toulouse, France}

\author{R. J. \surname{Nicholas}}
 \email{r.nicholas@physics.ox.ac.uk}
 \affiliation{Clarendon Laboratory, University of Oxford, Parks Road, Oxford, OX1 3PU, UK}

\date{\today}

\begin{abstract}
We have investigated the band structure at the $\Gamma$ point of the three-dimensional (3D) topological insulator Bi$_2$Se$_3$
using magneto-spectroscopy over a wide range of energies ($0.55-2.2$\,eV) and in ultrahigh magnetic fields up to 150\,T. At such
high energies ($E>0.6$\,eV) the parabolic approximation for the massive Dirac fermions breaks down and the Landau level
dispersion becomes nonlinear. At even higher energies around 0.99 and 1.6 eV, new additional strong absorptions are observed with
a temperature and magnetic-field dependence which suggest that they originate from higher band gaps. Spin orbit splittings for
the further lying conduction and valence bands are found to be 0.196 and 0.264 eV.
\end{abstract}

\maketitle

Topological insulators, which have a finite insulating bandgap in the bulk form, but which can become gapless at the surface,
have been intensively investigated as a new class of quantum matter.~\cite{Hasan10, Qi11, Ando13} Their surface states have
topologically-protected Dirac fermions with spins locked to their translational momentum due to a strong spin-orbit coupling,
giving them considerable potential to be used in future electronic and spintronic devices.~\cite{Mellnik14, Kondou16} The
experimental observation of a single Dirac cone in the surface state by angle-resolved photoemission spectroscopy (ARPES) was
reported for Bi$_2$Se$_3$,~\cite{Xia09,Hsieh09} Bi$_2$Te$_3$~\cite{Chen09} and Sb$_2$Te$_3$,~\cite{Hsieh09a} confirming that they
are three-dimensional (3D) topological insulators with large bandgap in the bulk (0.1--0.3 eV). This observation stimulated a
large body of work to investigate the unconventional properties arising from the topological surface states. However, there have
been few experimental studies of their band structures in the bulk, despite the fact that the origin of topological surface
states is critically dependent on the band inversion in the bulk crystal. In Bi$_2$Se$_3$, the bulk band structure has been
described using a massive Dirac Hamiltonian with a negative mass parameter caused by band inversion due to strong spin-orbit
coupling.~\cite{Zhang09, Liu10} While ARPES studies have reported the existence of a camelback-like band structure at the
$\Gamma$ point ``near the surface'' of Bi$_2$Se$_3$ consistent with the band inversion in the bulk,~\cite{Hsieh09} other
techniques such as magneto-transport and infrared spectroscopy favor a bulk behavior with a direct bandgap at the $\Gamma$ point
(\emph{i.e.} no camelback-like structure).~\cite{Post13, Piot16} The latter results are supported by theoretical $GW$
calculations~\cite{Yazyev12, Aguilera13} which suggest that electron-electron interactions suppress the camelback-like structure
in the bulk bands of Bi$_2$Se$_3$.

Magneto-optical studies recently revealed the band character in the bulk of Bi$_2$Se$_3$, which has been described by a massive
Dirac Hamiltonian with a negative mass term.~\cite{Orlita15} Remarkably, the observed evolution of the Landau-level energies in
Bi$_2$Se$_3$ remain linear even up to 32\,T, indicating that for the energies investigated ($E<0.6$\,eV), the band dispersion in
the bulk of Bi$_2$Se$_3$ is almost perfectly parabolic. The $GW$ calculations~\cite{Yazyev12, Aguilera13} have suggested the
existence of higher direct bandgaps at the $\Gamma$ point, in the near infrared and visible range. The higher conduction and
valence band can influence carriers in the lower bands, especially in the presence of an applied magnetic field. To date only
limited evidence of the higher band transitions has been presented~\cite{Eddrief16} from fitting of the dielectric function at
room temperature.

In this paper, we present magneto-optical studies of single crystal thin-film Bi$_2$Se$_3$ using ultrahigh magnetic fields up to
150\,T over a wide range of energies (0.55 to 2.2\,eV). The single crystal thin films are grown  on $c$-plane sapphire substrates
using molecular beam epitaxy, as described in detail in Refs.\onlinecite{Collins14,Collins15}. We reveal the nonlinear
magnetic-field evolution of the Landau levels of fundamental bands in the bulk form of Bi$_2$Se$_3$, which is explained by the
massive Dirac model with the negative mass term. At high energies, $\sim$0.99 and 1.6 eV, additional strong absorptions are
observed. From their temperature and magnetic-field dependence, they are assigned as the lowest interband Landau level
transitions of 2$^{nd}$ and 3$^{rd}$ bandgaps, respectively.

\begin{figure}
    \centering
    \hfill
    \includegraphics[width=1.0\columnwidth]{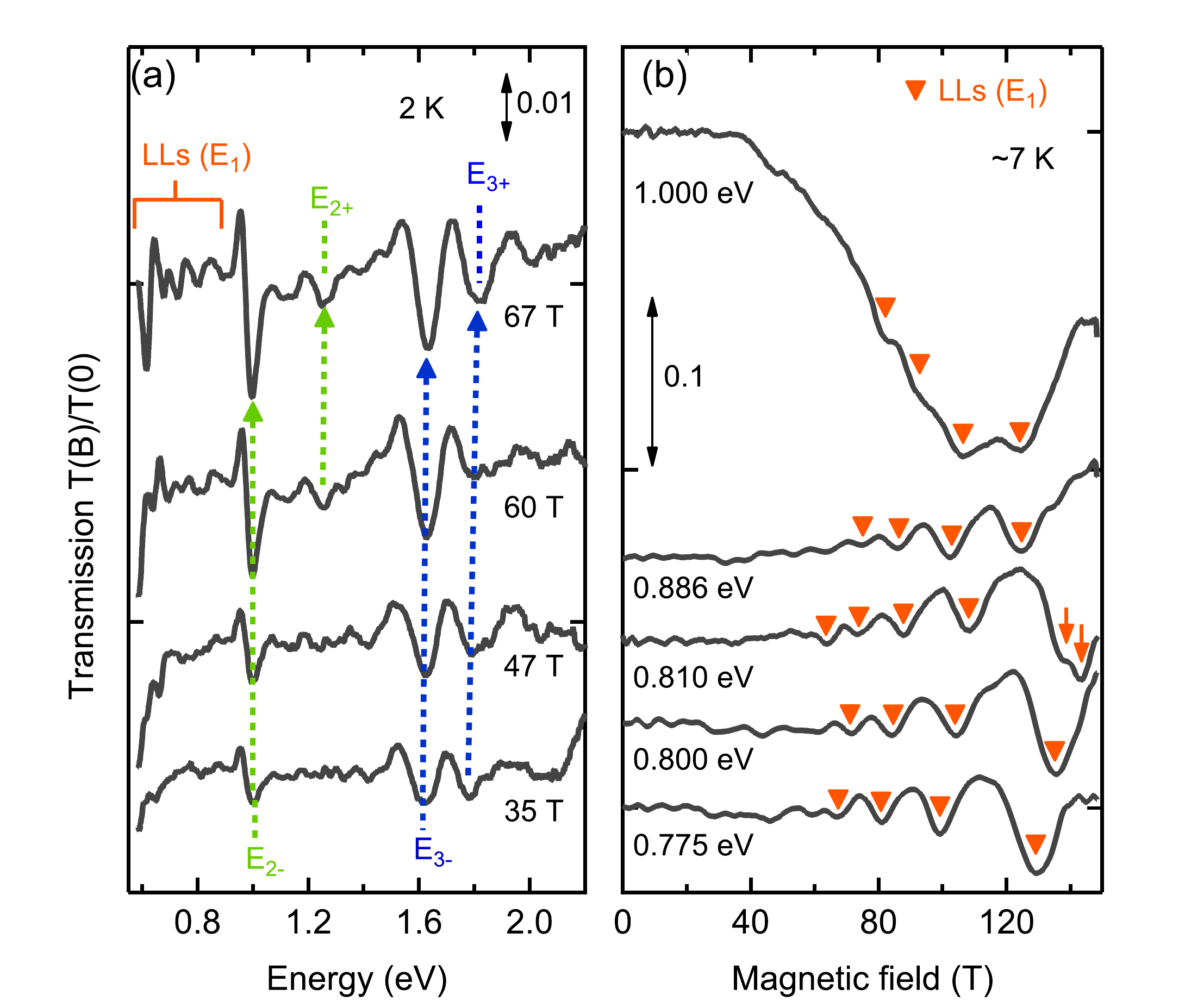}
    \hfill\null
\caption{(a) Low-temperature differential transmission spectra of Bi$_2$Se$_3$ at different magnetic fields of 35,
47, 60 and 67 T. (b) Magneto transmission of Bi$_2$Se$_3$ at different energies at 7 K. (a, b) Spectra were shifted vertically for
clarity. }\label{fig1}
\end{figure}

Figure\,\ref{fig1}(a) shows representative differential (normalized by dividing by zero-field spectra) transmission spectra,
$T(B)/T(0)$, in Bi$_2$Se$_3$ taken at 35, 47, 60 and 67\,T which reveal a number of well resolved absorption minima. At energies
below 0.95\,eV, we observe features due to inter Landau level absorption across the direct gap (transitions labeled $E_1$) at the
$\Gamma$-point. To study these absorptions in detail we performed magnetic field dependent measurements up to 150\,T, using an
explosive single turn coil and single frequency sources~.\cite{Nicholas13} Figure\,\ref{fig1}(b) shows normalized
magneto-transmission, $T(B)/T(0)$, of Bi$_2$Se$_3$ for excitation energies of 0.775, 0.800, 0.810 and 0.886 eV around $T=7$\,K up
to 150\,T. The transmission minima denoted by triangular symbols in Fig.\,\ref{fig1}(b) show clear shifts to higher magnetic
fields with increasing energies, which correspond to the interband Landau level transitions of the fundamental band gap
($\sim$0.2 eV) in the bulk of Bi$_2$Se$_3$~\cite{Orlita15}.

\begin{figure}
    \centering
    \hfill
    \includegraphics[width=1.0\columnwidth]{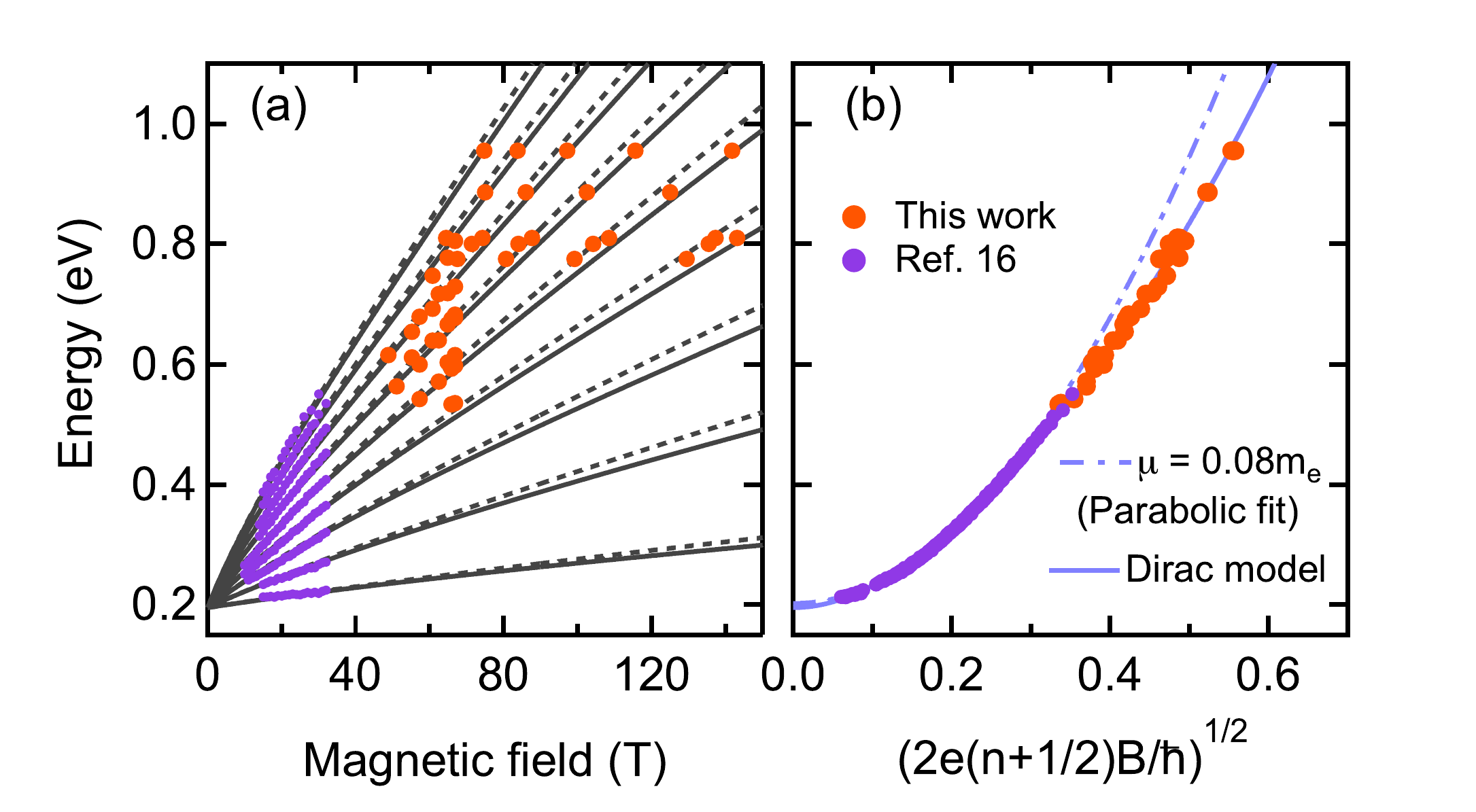}
    \hfill\null
\caption{(Color online) (a)Low-temperature interband Landau level fan chart for 1st (fundamental) bandgap in Bi$_2$Se$_3$. Dashed
and solid lines are interband Landau levels obtained by the 4$ \times $4 massive Dirac Hamiltonian with the electron-hole
asymmetry. (b)Energy-momentum dispersion of Bi$_2$Se$_3$. Broken line shows the fitting for parabolic dispersion. Inset shows the
deviation from the parabolic dispersion. }\label{fig2}
\end{figure}

Figure\,\ref{fig2}(a) shows the Landau level fan chart combining our transmission data with the low-energy data of
Orlita~\textit{et al.}.~\cite{Orlita15} The dipole allowed ($\Delta n = \pm 1$) interband inter Landau levels transitions,
calculated using the $4 \times 4$ massive Dirac Hamiltonian~\cite{Zhang09, Liu10} reproduce the data perfectly, with a band gap
of 2$\Delta$=0.19~eV, Fermi velocity $v_D=(0.465 \pm 0.05) \times 10^{6}$ ms$^{-1}$, negative mass term $M=-(17 \pm
0.5)$\,eV\AA$^2$. The electron-hole asymmetry parameter $C=(3 \pm 0.5)$ eV\AA$^2$, which reflects the different electron and hole
effective masses, was determined from the observed splitting of interband Landau level transition, indicated by arrows in
Fig.\,\ref{fig1}(b).

Our fitting parameters are identical within experimental error with those obtained by Orlita~\textit{et al.},~\cite{Orlita15}
with the important exception of the negative mass term. Our higher energy data places stronger constraints on the value of $M$
and our value of $|M|=17$\,eV\AA$^2$ is roughly 25\% smaller than found by fitting to the low energy data of Orlita~\textit{et
al.}. This implies that the condition $\hbar^2v_D^2$=$-4M\Delta$ required in the Dirac model to have a strictly parabolic
dispersion over a wide energy range, is not fulfilled as well as previously thought. We have $\hbar^2 v_D^2 \simeq
9.6$\,eV$^2$\AA$^2$ but with a significantly smaller value for $-4M\Delta \simeq 6.8$\,eV$^2$\AA$^2$. This also has implications
for the magnetic field $B_c = \hbar\Delta/|eM|$ at which the $n=0$ Landau levels of the conduction and valence bands cross. The
lower value of $|M|$ found here shifts $B_c$ to higher magnetic field ($B_c \simeq 390$\,T compared to the value of around
$300$\,T estimated by Orlita~\textit{et al.}).

In Fig.\,\ref{fig2}(b), the experimentally obtained inter band Landau level transition energies, with different indexes are
scaled into the same energy-momentum dispersion of Bi$_2$Se$_3$, deduced from the relation between the momentum $k$ and magnetic
field $B$. Neglecting electron hole asymmetry, and assuming a parabolic dispersion, the energy of a dipole allowed transition $n
\rightarrow n+1$ or $n+1 \rightarrow n$ is given by $E_n = 2\Delta + [(n+\gamma + 1) + (n + \gamma)]\hbar\omega_c$ with
$\gamma=0$ for massive Dirac particles. This can be rewritten using the reduced exciton mass $\mu = m^*/2$ to give $E_n = 2\Delta
+ (n + 1/2)eB/\mu$. Equating the magnetic energy to $\hbar^2k^2/2\mu$ gives an expression for the $k$ vector $k = \sqrt{2eB(n +
1/2)/\hbar}$, which is equivalent to the Bohr-Sommerfeld quantization of area in \textit{k}-space which is independent of the
particle mass values assumed. In the plot of the observed transition energies versus $k = \sqrt{2eB(n + 1/2)/\hbar}$, all the
data collapse onto a single curve. The broken line in Fig.\,\ref{fig2}(b) shows a parabolic dispersion using a band gap
2$\Delta=0.2$\,eV and reduced exciton mass $\mu$=0.08$m_0$ ($m_0$ is the free electron mass). For energies above 0.6\,eV a clear
deviation from the simple parabolic model is observed reflecting the transition towards a $\sqrt{B}$ dependence for the energy of
Dirac fermions, which is clearly visible in Fig.\,\ref{fig2} (a). The reduced mass is consistent with recent results obtained by
Shubnikov de Haas (SdH) measurements for $n$-type and $p$-type samples of Bi$_2$Se$_3$; the range of effective masses of the
electrons and holes respectively is 0.12--0.16\,$m_0$ and 0.23--0.25\,$m_0$ in the literature,~\cite{Piot16, Eto10, Analytis10}
corresponding to a value for the reduced mass, $\mu$, in the range 0.079--0.098 $m_0$. It is also possible to estimate the
effective masses from the Fermi velocity $v_D$ and the electron-hole asymmetry term $C$ using
$m_e=2\hbar^2/(\hbar^2v_D^2/\Delta+4C)$  and $m_h=2\hbar^2/(\hbar^2v_D^2/\Delta-4C)$, giving $m_e=0.138\,m_0$ and
$m_h=0.176\,m_0$, respectively.

\begin{figure}
    \centering
    \hfill
    \includegraphics[width=1.0\columnwidth]{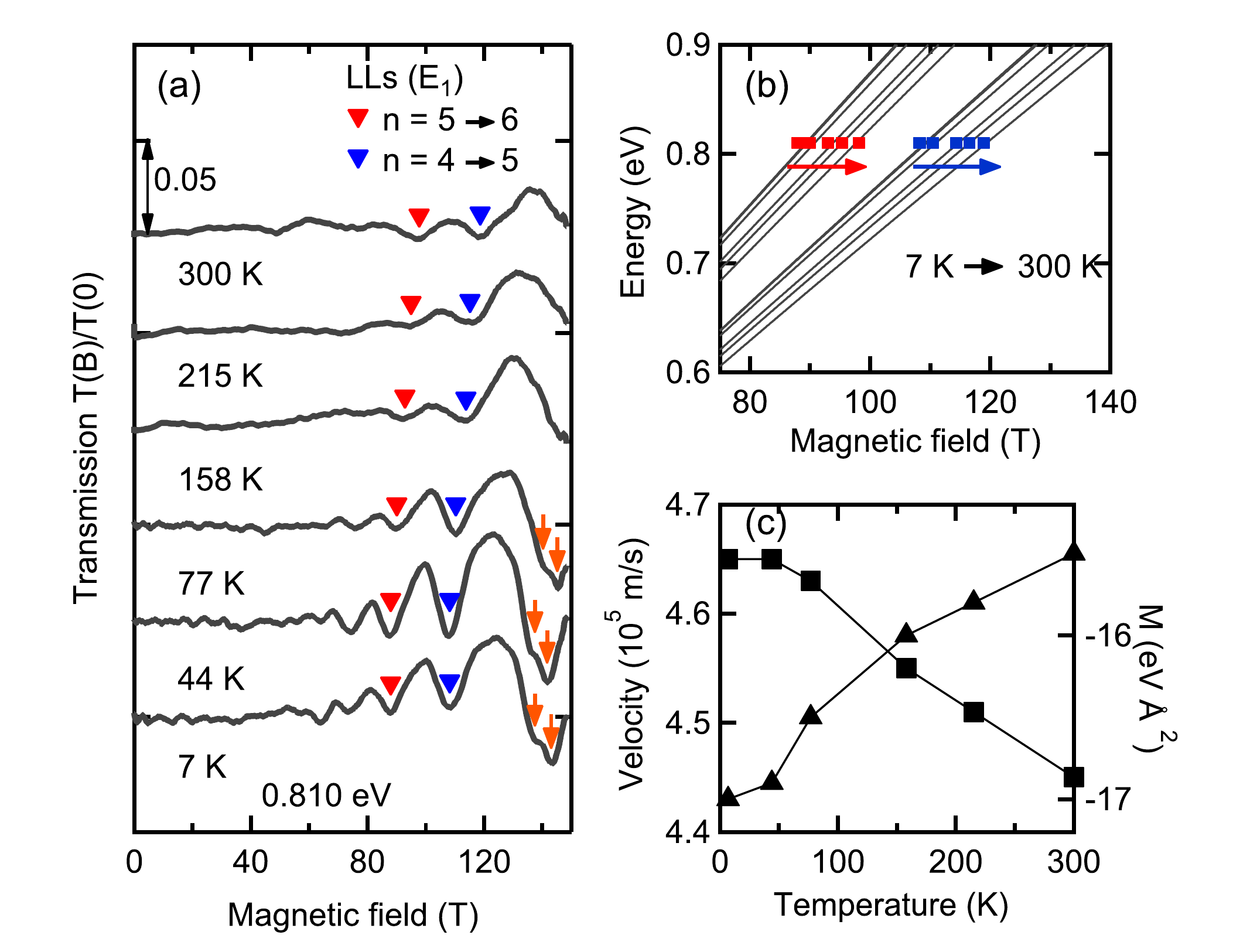}
    \hfill\null
\caption{(a) Magneto-transmittance of Bi$_2$Se$_3$ at the energy of 0.810\,eV measured for different temperatures
in fields up to 150 T. The position of the inter Landau level transitions are indicated by triangles. The splitting due to
electron-hole asymmetry is indicated by arrows. The spectra have been shifted vertically for clarity. (b) Energy of transitions
at 7\,K and 300\,K (symbols). The solid lines are fits to the Dirac like Hamiltonian. (c) Extracted Fermi velocity (squares) and
negative mass parameter (triangles) versus temperature. }\label{fig3}
\end{figure}

Figure\,\ref{fig3}(a) shows the magneto-transmission of Bi$_2$Se$_3$ for temperatures from 7 to 300 K with an excitation energy
of 0.810\,eV. The inter band Landau level transitions of the fundamental band gap show a clear shift to higher magnetic fields
with increasing temperatures. The observed shift cannot be attributed to a temperature dependence of the fundamental band gap
energy 2$\Delta$ in the massive Dirac model as the avoided band crossing at low energies has almost no influence on
the high energy inter band Landau level transitions. Instead we argue that this is due to a temperature dependence of the Fermi
velocity and the negative mass parameter. In Fig.\,\ref{fig3}(b) we plot the observed transition energies at 7 and 300\,K
together with the transition energies calculated using the Dirac like Hamiltonian. From these we deduce, in Fig.\,\ref{fig3}(c),
the temperature dependence of $v_D$ and $M$. By 300\,K the Fermi velocity has dropped to $v_D = 0.445 \times $10$^{6}$
m\,s$^{-1}$, significantly ($\simeq 6\%$) lower than the low temperature value. At the same time the negative mass parameter has
changed by around 10\% to $M \simeq -15.5$\,eV\AA$^2$ at 300K.

\begin{figure}
    \centering
    \hfill
    \includegraphics[width=1.0\columnwidth]{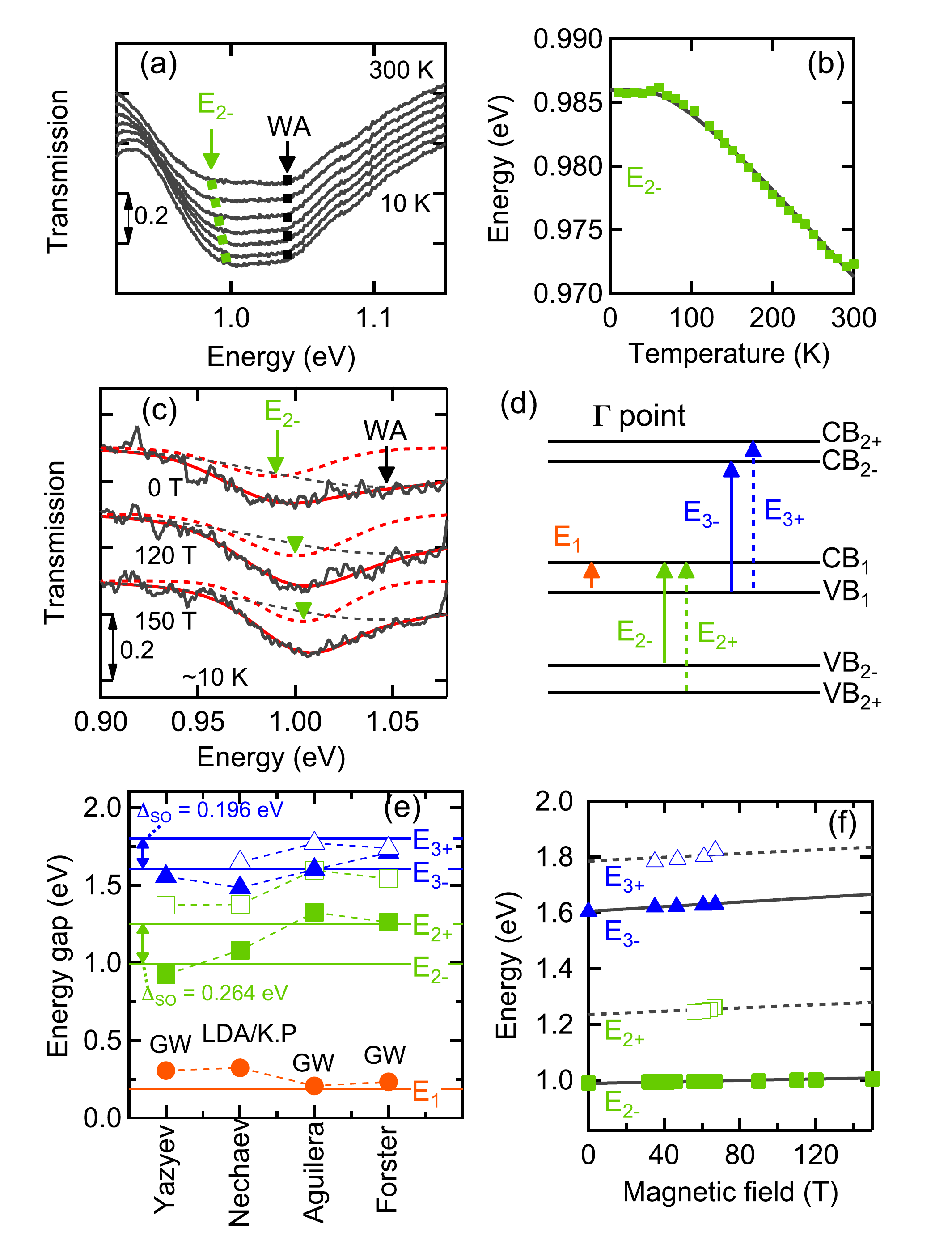}
    \hfill\null
\caption{(a) Transmission spectra of Bi$_2$Se$_3$ around 1.0 eV at 0 T at different temperatures of 10 and 50--300 K (50 K step).
(b) Temperature dependence of 2$^{nd}$ bandgap energy of Bi$_2$Se$_3$. Solid line is fitting by the expression in the text. (c)
Low-temperature magneto-transmission spectra of Bi$_2$Se$_3$ at different magnetic fields. Spectra in (a) and (c) were shifted
vertically for clarity and the weak feature labelled WA is due to water vapour absorption. (d) Schematic picture of band
structure in the bulk of Bi$_2$Se$_3$ around the $\Gamma$ point. (e) Low-temperature Landau fan chart of Bi$_2$Se$_3$ for
2$^{nd}$ and 3$^{rd}$ bandgaps and split-off bandgaps. Solid lines show the interband Landau levels obtained by the model
described in the text.}\label{fig4}
\end{figure}

We now turn our attention to the transitions observed at higher energies in Fig.\,\ref{fig1}(a). There is clear evidence for a
2$^{nd}$ band gap, with a pair of absorptions at 0.986\,eV and 1.25\,eV which we assign to a pair of transitions from a lower
spin orbit split valance band~\cite{Liu10} to the lowest empty conduction band (labeled $E_{2\pm}$). At still higher energies, we
observe absorption from a second pair of transitions for the 3$^{rd}$ band gap at 1.604\,eV and 1.8\,eV due to transitions
(labeled $E_{3\pm}$) from the lowest occupied valence band to a higher spin orbit split conductance band. The observed optical
transitions at the $\Gamma$ point are shown schematically in Fig.\,\ref{fig4}(d). The transition energies (solid lines) are
compared with the transition energies predicted by several LDA/GW band structure calculations
\cite{Yazyev12,Nechaev16,Aguilera13,Foerster15} (symbols) shown in Fig.\,\ref{fig4}(e). The absolute values of the band gaps are
in generally good agreement with theory and the spin-orbit splittings of the lower valence band (0.264 eV) and upper conduction
band (0.196 eV) are also in good agreement with the predicted values. These higher band gaps are also consistent with those (1.0
eV and 1.6 eV) estimated from fitting the dielectric functions of bulk Bi$_2$Se$_3$ at 300\, K as reported in
Ref.~\onlinecite{Eddrief16}, although the spin-orbit splitting could not be resolved.

We now examine the temperature and magnetic-field dependence of the higher-energy absorptions. Figure\,\ref{fig4} (a) shows the
temperature dependence of the transmission spectra at temperatures of 10\,K and 50 to 300\,K (50\,K step). The $E_{2-}$
transition around 0.99 eV from the lower spin orbit split valence band (2$^{nd}$ band gap) shows a clear shift towards higher
energies with decreasing temperature, which is plotted in Fig.\,\ref{fig4}(b) (the small feature labelled WA is a water
absorption around 1.04\,eV.~\cite{Collins25}) The temperature dependence was fitted using the well known expression for the
temperature dependence of a semiconductor bandgap,~\cite{ODonnell91}
\begin{eqnarray}
    \label{eqn:H}
     E_g (T)=E_g(0)-S\langle\hbar\omega\rangle\left(\coth{(\frac{\langle\hbar\omega\rangle}{2k_BT})}-1\right),
\end{eqnarray}
where $E_g(T)$ is the band gap at the temperature $T$ with $E_g(0) = 0.986$\,eV, $S=0.423$ is a dimensionless coupling constant
and $\langle\hbar\omega\rangle = 19.4$\,meV is the average phonon energy (see Fig. S3 for the temperature dependence of 1.6\,eV
absorption). The average phonon energy is within the range of optical phonon energies (4.6--21.6 meV) seen in Raman spectroscopy
in the literature.~\cite{Zhang11}

Figure\,\ref{fig4}(c) shows the magnetic field dependence of the transmission spectra, $T(B)$, at 0, 120 and 150 T, measured
using a super-continuum light source ("white-light laser") in the range 1.1 to 1.7\,µm. A single 20ns light pulse is synchronized
with the magnetic field pulse from the single turn coil and the spectrometer acquisition window. The light pulse is beamsplit and
the sample transmission normalised against the same pulse to account for pulse to pulse variability. The $E_{2-}$ around 0.99\,eV
at 0\,T shows a small but clear shift towards higher energy with increasing magnetic fields ($\sim$15 meV shift at 150\,T) and a
field induced increase in intensity.  This small diamagnetic shift compared with that of $\sim$100 meV (Fig. \ref{fig2}) seen for
the fundamental band gap is consistent with the assignment of this transition as a strongly bound excitonic state associated with
the second band gap. The increase in intensity and shift produced by the magnetic field also result in the strong differential
transmission signals seen in Fig. \ref{fig1} for the $E_{2}^{\pm}$ and $E_{3}^{\pm}$ transitions, which allows them to be
distinguished from the higher quantum number interband Landau level transitions of band gap $E_1$ seen in a similar energy range
in Fig.\,\ref{fig1}(a).

The relative strenghts of the band edge $E_{2-}$ transition and the higher landau level $E_{1}$ transitions can be seen in Fig.
\ref{fig1}(b) from the magnetic field dependent transmission in the 100T range of magnetic fields at 1.0\,eV where changes as
large as 20\% can be seen, compared to the few \% changes for the higher Landau levels. This is further evidence that the
$E_{2-}$ and $E_{3-}$ transitions are not high-order Landau levels of the fundamental band gap, but the lowest interband Landau
level transitions of the 2$^{nd}$ and 3$^{rd}$ band gaps with intensities enhanced by excitonic interactions. Such behavior is
typical of that seen, for example, in magneto-transmission studies in the 100T range of magnetic fields for excitonic states in
GaSe~\cite{Watanabe03} and the organic lead halide perovskite CH$_3$NH$_3$PbI$_3$.~\cite{Miyata15}

The exciton diamagnetic shift can be used to estimate some properties of the higher band gap excitonic states using the
expression,~\cite{Miura08, Walck98}
\begin{eqnarray}
    \label{eqn:Diamag}
     \Delta E_g(B)=\frac{e^2\langle r^{2}\rangle B^{2}}{8\mu},
\end{eqnarray}
which does not require a knowledge of the dielectric constant and where $r$ is the size of the exciton wavefunction and $\mu$ is
the exciton reduced effective mass between the $CB_1$ and $VB_{2-}$ interband transitions. Recent recent $\bf {k.p}$
calculations~\cite{Nechaev16} suggest that the effective masses in the lower valuence band and upper conduction bands will be
comparable to those for $CB_1$ and $VB_1$, allowing us to approximate $\mu$ to be on the order of $\mu$=0.1\,$m_0$. This means
that Eq.\ref{eqn:Diamag} predicts that the exciton size is only on the order of $\sqrt{\langle r^{2}\rangle}$= 1.8 nm.  This
seems surprisingly small given the well know high values for the dielectric constant of Bi$_2$Se$_3$, which has been shown to be
of order 30 even up to photon energies as high as 2 eV.~\cite{Eddrief16}  However such a value would be consistent with the idea
that there is be a significant excitonic contribution to these higher band transitions which influences the band edge absorption.

In conclusion we have used magnetospectroscopy over a wide range of energies (0.55 to 2.2 eV) and magnetic fields up to 150 T to
investigate the bulk band structure of the topological insulator Bi$_2$Se$_3$. The interband Landau level transitions of the
fundamental bandgap are observed to deviate from the previously reported~\cite{Orlita15} linearity at high-energies, but can be
described by a 4$ \times $4 massive Dirac Hamiltonian with a negative mass term arising from band inversion caused by strong
spin-orbit coupling. Furthermore, in high magnetic fields the differential absorption spectra, $T(B)/T(0)$,  reveal new strong
resonances which were assigned as the lowest interband Landau level transitions of the 2$^{nd}$ and 3$^{rd}$ bandgaps observed at
high energies around 0.99 and 1.6 eV, with clearly resolved spin-orbit splittings. From their temperature and magnetic-field
dependence, they were assigned as excitonic band edge transitions for these higher bandgaps.

\section{Acknowledgements} This work was supported by EPSRC (UK) via its membership of the EMFL (grant no.\ EP/N01085X/1), the TERASPEC project of the Emergence program of IDEX Toulouse, and MEGATER project funded by NEXT Toulouse, by ANR JCJC project milliPICS, the R{\'e}gion Midi-Pyr\'en\'ees under contract MESR 13053031,
the BLAPHENE and STRABOT projects under IDEX program Emergence, and by ``Programme Investissements d'Avenir'' under the program
ANR-11-IDEX-0002-02, reference ANR-10-LABX-0037-NEXT. This work arises from research funded by the John Fell Oxford University Press Research Fund. LBD acknowledges financial support from EPSRC and the Science and Technology Facilities Council (UK).



%

\end{document}